\documentclass[prl,twocolumn,showpacs,floatfix]{revtex4}

\usepackage{amsmath}
\usepackage{graphicx}
\usepackage{amsfonts}
\usepackage{amssymb}

\begin{document}

\title{Completely Positive Post-Markovian Master Equation via a Measurement
Approach}
\author{A. Shabani$^{(1)}$ and D.A. Lidar$^{(2)}$}
\affiliation{$^{(1)}$Physics Department, University of Toronto, 60 St. George St.,
Toronto, Ontario M5S 1A7, Canada$^{(2)}$Chemical Physics Theory Group,
Chemistry Department, and Center for Quantum Information and Quantum
Control, University of Toronto, 80 St. George St., Toronto,
Ontario M5S 3H6, Canada}

\begin{abstract}
A new post-Markovian quantum master equation is derived, that includes bath
memory effects via a phenomenologically introduced memory kernel $k(t)$. The
derivation uses as a formal tool a probabilistic single-shot
bath-measurement process performed during the coupled system-bath evolution.
The resulting analytically solvable master equation interpolates between the
exact Nakajima-Zwanzig equation and the Markovian Lindblad equation. A
necessary and sufficient condition for complete positivity in terms of
properties of $k(t)$ is presented, in addition to a prescription for the
experimental determination of $k(t)$. The formalism is illustrated with
examples.
\end{abstract}

\maketitle

An open quantum system is one that is coupled to an external environment 
\cite{Breuer:book,Alicki:87}. Such systems are of fundamental interest, as
the notion of a closed system is always an idealization and approximation.
Open quantum systems tend to decohere, and for this reason have recently
received intense consideration in quantum information science, where
decoherence is viewed as fundamental obstacle to the construction of quantum
information processors \cite{Nielsen:book}. It is possible to write down an
exact dynamical equation for an open system, but the result -- an
integro-differential equation \cite{Nakajima:58Zwanzig:60a} -- is mostly of
formal interest, as such an exact equation can almost never be solved
analytically or even numerically. In contrast, when one makes the Markovian
approximation, i.e., when one neglects all bath memory effects, the
resulting Lindblad master equation \cite{Gorini:76Lindblad:76,Alicki:87} is
formally solvable and amenable to numerical treatment. Moreover, the
desirable property of complete positivity \cite{Kraus:83} is maintained
(see, however, \cite{Pechukas:94Pechukas+Alicki:95} for a debate on the
importance of this property). A coveted goal of the theory of open quantum
systems \cite{Breuer:book,Alicki:87} is a \textquotedblleft
post-Markovian\textquotedblright\ master equation that \emph{(i) generalizes
the Markovian Lindblad equation so as to include bath memory effects, at the
same time (ii) remains both analytically and numerically tractable, and
(iii) retains complete positivity}. A variety of post-Markovian master
equations have been proposed and analyzed, e.g., \cite%
{Breuer:book,Shibata:77Chaturvedi:79,Imamoglu:94,Royer:96Royer:03,Garraway:97Breuer:99knezevic:012104,gambetta:012108breuer:022115,Strunz:99Yu:2000,barnett:033808,Daffer:03,Breuer:04}%
. However, one of the desirable properties (i)-(iii) above is typically
lost: e.g., in the case of time-convolutionless master equations (e.g., \cite%
{Royer:96Royer:03}) one may lose complete positivity, while in the case of
nonlocal stochastic Schrodinger equations (e.g., \cite{Strunz:99Yu:2000})
one loses analytical solvability. In this work we propose a new
post-Markovian master equation that satisfies all of the desirable
properties (i)-(iii) above. The key idea we introduce is an interpolation
between the generalized measurement interpretation of the exact Kraus
operator sum map \cite{Kraus:83}, and the continuous measurement
interpretation of Markovian-limit dynamics \cite%
{Dalibard:92Gisin:92Plenio:98,Breuer:04}.

\textit{Review of quantum measurements approach to open system dynamics}.---
Consider a quantum system $S$ coupled to a bath $B$ (with respective Hilbert
spaces $\mathcal{H}_{S},\mathcal{H}_{B}$), evolving unitarily under the
total system-bath Hamiltonian $H_{SB}$. The exact system dynamics is given
by tracing over the bath degrees of freedom \cite%
{Breuer:book,Alicki:87,Nielsen:book} 
\begin{equation}
\rho(t)=\mathrm{Tr}_{B}[U(t)\rho _{SB}(0)U^{\dag }(t)],  \label{system}
\end{equation}
where $\rho(t)$ is the system state, $\rho _{SB}(0)=\rho(0)\otimes \rho
_{B}(0)$ is the initially uncorrelated system-bath state, and $U(t)=\mathcal{%
T}\mathsf{\exp }(-i\int_{0}^{t}H_{SB}(t^{\prime })dt^{\prime })$ ($\mathcal{T%
}$ denotes time-ordering; we set $\hbar =1$ and for simplicity work in the
interaction picture with respect to both system and bath). Eq.~(\ref{system}%
) can be rewritten in terms of an operator sum (the Kraus representation 
\cite{Kraus:83}) 
\begin{equation}
\rho(t)=\sum_{k}A_{k}^{\dag }(t)\rho(0)A_{k}(t),  \label{eq:Kraus}
\end{equation}
where $\mathrm{Tr}[\rho(t)]=1\Leftrightarrow \sum_{k}A_{k}(t)A_{k}^{\dag
}(t)=I$.

\begin{figure}[tbp]
\hspace{1.5cm} \includegraphics[width=8.5cm]{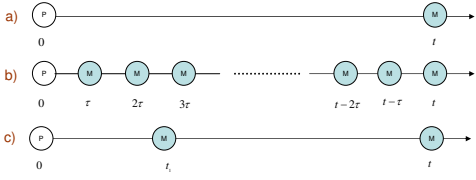}  \vspace{-4cm}
\caption{Measurement approach to open system dynamics. P$=$preparation, M$=$
measurement, time proceeds from left to right. a) Exact Kraus operator sum
representation, b) Markovian approximation, c) Single-shot measurement, d)
Single-shot measurement followed by Markovian dynamics. }
\label{fig1}
\end{figure}

Let us now recall how to derive the \emph{exact} Eq.~(\ref{system}) from a
measurement picture [Fig.~\ref{fig1}a]. Imagine the bath acting as a probe
coupled to the system at $t=0$, with the interaction given by $H_{SB}$ as
above. To study the state of the system a \emph{single} projective
measurement is performed on the \emph{bath} at time $t$, with a complete set
of projection operators $|i\rangle \langle i|$, $\mathcal{H} _{B}=\mathrm{%
Span}\{|i\rangle \}$. The measurement yields the result $k$ and collapses
the state of the bath to the corresponding eigenstate $|k\rangle $. This
happens with probability $p_{k}=\mathrm{Tr}_{S}[\langle k|\rho
_{SB}(t)|k\rangle ]$, and the system density matrix reduces to $%
\rho^{k}(t)=\langle k|\rho _{SB}(t)|k\rangle /p_{k}=:A_{k}^{\dag
}\rho(0)A_{k}/p_{k}$, where $A_{k} $ are the Kraus operators. If we repeat
this process for an identical ensemble initially prepared in state $\rho
_{SB}(0)$ the average system density matrix becomes $\rho(t)=\sum_{k}p_{k}%
\rho^{k}(t)=\mathrm{Tr}_{B}[U(t)\rho _{SB}(0)U^{\dag }(t)]$, which is just
Eq.~(\ref{system}), thus affirming the validity of this bath-measurement
interpretation of open system dynamics. The corresponding map $\Phi$ is
completely positive (CP) \cite{CP}.

In contrast, in the Markovian limit the most general CP system dynamics is
given in the interaction picture by the Lindblad equation \cite%
{Gorini:76Lindblad:76}

\begin{equation}
\frac{\partial \rho}{\partial t}=\mathcal{L}\rho :=-\frac{1}{2}
\sum_{a}a_{\alpha }([F_{\alpha },\rho F_{\alpha }^{\dag }]+[F_{\alpha
}\rho,F_{\alpha }^{\dag }]).  \label{eq:Lind}
\end{equation}
The Lindblad operators $F_{\alpha }$'s are bounded operators acting on $%
\mathcal{H_{S}}$, and the $a_{\alpha } \ge 0$ are constants that describe
decoherence rates. Now let us recall how also the Lindblad equation can be
given a measurement interpretation. Expanding Eq.~(\ref{eq:Lind}) to first
order in the short time interval $\tau $ yields $\rho(t+\tau )=(I-\frac{\tau 
}{2}\sum_{\alpha }F_{\alpha }^{\dag }F_{\alpha })\rho(t)(I-\frac{\tau }{2}%
\sum_{\alpha }F_{\alpha }^{\dag }F_{\alpha }) + \tau \sum_{\alpha }F_{\alpha
}\rho(t)F_{\alpha }^{\dag }$. To the same order we also have the
normalization condition $(I-\frac{\tau }{2 }\sum_{\alpha }F_{\alpha }^{\dag
}F_{\alpha })(I-\frac{\tau }{2}\sum_{\alpha }F_{\alpha }^{\dag }F_{\alpha
})+\tau \sum_{\alpha }F_{\alpha }^{\dag }F_{\alpha }=I$. Thus the Lindblad
equation has been recast as a Kraus operator sum (\ref{eq:Kraus}), but only
to first order in $\tau $, the coarse-graining time scale for which the
Markovian approximation is valid \cite{Lidar:CP01}. Clearly, then, we again
have a measurement interpretation, wherein as before the bath functions as a
probe coupled to the system while being subjected to a \emph{continuous}
series of measurements at each infinitesimal time interval $\tau $ [Fig.~\ref%
{fig1}b)]. This is the well-known quantum jump process \cite%
{Dalibard:92Gisin:92Plenio:98}, wherein the measurement operators are $I-%
\frac{\tau }{2}\sum_{\beta }F_{\beta }^{\dag }F_{\beta }$ (the
``conditional'' evolution) and $\sqrt{\tau }F_{\alpha }$ (the ``jump'').

We have thus seen how a measurement picture leads to the two limits of exact
dynamics (via an evolution of the coupled system-bath followed by a single
generalized measurement at time $t$), and Markovian dynamics (via a series
of measurements interrupting the joint evolution after each time interval $%
\tau $). With this in mind it is now easy to see that by relaxing the
many-measurements process one is led to a less restricted approximation than
the Markovian one. Here we use this observation to derive a post-Markovian
master equation based on a probabilistic single-shot measurement process.

\textit{Derivation of a post-Markovian master equation}.--- The first stage
of exerting an approximation on the exact Eq.~(\ref{system}) should be to
include one extra measurement in the time interval $[0,t]$. Thus we consider
the following process: a probe (bath) is coupled to the system at $t=0$,
they evolve jointly for a time $t^{\prime }$ ($0\leq t^{\prime }<t$) such
that at $t^{\prime }$ the system state is $\Lambda (t^{\prime })\rho (0)$,
where $\Lambda (t^{\prime })$ is a one-parameter map, at which moment the
extra generalized measurement is performed on the bath. $\Lambda $ does not
depend on $t$ since the bath resets upon measurement. System and bath
continue their coupled evolution between $t^{\prime }$ and $t$, upon which
the final measurement is applied. This is illustrated in Fig.~\ref{fig1}c.
Since this intermediate measurement determines the system state $|\psi
\rangle $ at $t^{\prime }$, after time $t-t^{\prime }$ the system state will
be $\rho (t)=\Lambda (t-t^{\prime })\rho (t^{\prime })$. It is important to
stress that $\rho (t^{\prime })$ can\emph{not} be written as $\Lambda
(t^{\prime })\rho (0)$, since the measurement selects $\rho (t^{\prime })$
at random.

The time $t^{\prime }$ characterizes bath memory effects and must be
determined as a function of time-scales characterizing the evolution. We do
this by introducing a bath memory function (kernel) $k(t-t^{\prime },t)$
that assigns weights to different measurements. To derive a master equation we discretize the time interval 
$[0,t]$ into $N$ equal segments of length $\epsilon $, and express
$t=N\epsilon,t^\prime = m\epsilon$. We then have the
weighted average $\rho (t=N\epsilon )=\sum_{m=1}^{N}k((N-m)\epsilon ,N\epsilon
)\Lambda ((N-m)\epsilon )\rho (m\epsilon ) = \sum_{m=1}^{N}k(m\epsilon ,N\epsilon
)\Lambda (m\epsilon )\rho ((N-m)\epsilon )$. From hereon we assume that $%
\Lambda $ is trace-preserving, whence $k$ must be normalized so that $%
\sum_{m=1}^{N}k(m\epsilon ,N\epsilon)=1$ ($k(t^{\prime },t)=0$ for $%
t^{\prime }\notin \lbrack 0,t]$), though an exception to this will arise
below. We then have (for $N\geq 1$) 
\begin{eqnarray}
&&\rho (N\epsilon )-\rho ((N-1)\epsilon )=\sum_{m=1}^{N-1}k(m\epsilon
,(N-1)\epsilon )\Lambda (m\epsilon )\times  \notag \\
&&\lbrack \rho ((N-m)\epsilon )-\rho ((N-m-1)\epsilon )]+\sum_{m=1}^{N-1}[k(m\epsilon ,N\epsilon ) \notag \\
&&-k(m\epsilon ,(N-1)\epsilon
)]\Lambda (m\epsilon )\rho ((N-m)\epsilon ) \notag \\
    &&+k(N\epsilon ,N\epsilon )\Lambda (N\epsilon )\rho (0).
    \label{Dif.}
\end{eqnarray}%
In order to arrive at a differential equation the term proportional to $%
\Lambda (N\epsilon )\rho (0)$ must be made to vanish. We therefore impose
the additional constraint $\lim_{\epsilon \rightarrow 0}k(N\epsilon
,N\epsilon )/\epsilon =0$. Taking the limits $\epsilon \rightarrow 0$, $%
m,N\rightarrow \infty $ such that $m\epsilon =t^{\prime }$ and $N\epsilon =t$%
, we convert the remaining terms in Eq.~(\ref{Dif.}) into differential form
by expressing $[\rho ((N-m)\epsilon )-\rho ((N-1-m)\epsilon )]/\epsilon
\rightarrow \frac{\partial \rho (t-t^{\prime })}{\partial (t-t^{\prime })}$
and $[k(m\epsilon ,N\epsilon )-k(m\epsilon ,(N-1)\epsilon )]/\epsilon
\rightarrow \frac{\partial k(t^{\prime },t)}{\partial t}$. Eq.~(\ref{Dif.})
then yields: $\frac{\partial \rho }{\partial t}=\int_{0}^{t}dt^{\prime
}[k(t^{\prime },t)\Lambda (t^{\prime })\frac{\partial \rho (t-t^{\prime })}{%
\partial (t-t^{\prime })}+\frac{\partial k(t^{\prime },t)}{\partial t}%
\Lambda (t^{\prime })\rho (t-t^{\prime })]$. We would like to arrive at a
proper integro-differential equation involving, on the right-hand-side
(RHS), only $\rho $ and not its derivative. We thus assume, \emph{only in
the derivative of }$\rho $ \emph{on the RHS}, that $\rho (t-t^{\prime
})=\Lambda (t-t^{\prime })\rho (0)$. Such an assumption is equivalent to the
standard procedure of first-order time-dependent perturbation theory, and
can, analogously, be iterated self-consistently to obtain higher-order
approximations. Expressing $\rho (0)=\Lambda ^{-1}(t-t^{\prime })\rho
(t-t^{\prime })$ we then obtain the post-Markovian dynamical equation%
\begin{eqnarray}
\frac{\partial \rho }{\partial t}&=&\int_{0}^{t}dt^{\prime }[k(t^{\prime
})\Lambda (t^{\prime })\dot{\Lambda}(t-t^{\prime })\Lambda ^{-1}(t-t^{\prime
  }) \notag \\
 && +\frac{\partial k(t^{\prime },t)}{\partial t}\Lambda (t^{\prime })]\rho
(t-t^{\prime }).
\label{equation-gen}
\end{eqnarray}
This new, formal master equation is the first main result of this work. Note
that in this integral form the constraint $\lim_{\epsilon \rightarrow
0}k(N\epsilon ,N\epsilon )/\epsilon =0$ can be lifted, as it cannot change
the value of the integral. 

To make further progress we now assume a Markovian form for the
superoperator: $\Lambda (t)=\exp (\mathcal{L}t)$. Here $\mathcal{L}$ can be interpreted as
the Lindblad generator [Eq.~(\ref{eq:Lind})]. Using this in Eq.~(\ref%
{equation-gen}) yields 
\begin{equation}
\frac{\partial \rho }{\partial t}=\int_{0}^{t}dt^{\prime }[k(t^{\prime },t)%
\mathcal{L}+\frac{\partial k(t^{\prime },t)}{\partial t}]\exp (\mathcal{L}%
t^{\prime })\rho (t-t^{\prime }).  \label{eq:main2}
\end{equation}%
This master equation is rather interesting and appears amenable to
analytical treatment, an undertaking which will be the subject of a future
study. To make even further progress, let us note that Eq.~(\ref{eq:main2})
automatically preserves $\mathrm{Tr}\rho $, even without requiring
normalization of $k$ via $\int_{0}^{t}k(t^{\prime },t)dt^{\prime }=1$. Since
the latter was needed above to ensure trace preservation, it can now be
dropped. This allows us to consider memory kernels satisfying $k(t^{\prime
},t)=k(t^{\prime })$. We thus arrive at our second main result: 
\begin{equation}
\frac{\partial \rho }{\partial t}=\mathcal{L}\int_{0}^{t}dt^{\prime
}k(t^{\prime })\exp (\mathcal{L}t^{\prime })\rho (t-t^{\prime })=\mathcal{L}%
k(t)\exp (\mathcal{L}t)\ast \rho (t),  \label{Equation}
\end{equation}%
where $\ast $ denotes convolution and $k$ no longer obeys any constraints.

Henceforth we confine our attention for simplicity and explicitness to the
new post-Markovian master equation~(\ref{Equation}), though some of the
results below are generalizable to Eq.~(\ref{equation-gen}). While $k$ is
still unspecified, we show below that it can be determined by an appropriate
quantum state tomography experiment. As we further show below, Eq.~(\ref%
{Equation}) satisfies all the conditions we stated in the introduction for a
\textquotedblleft desirable\textquotedblright\ post-Markovian master
equation. Finally, note that Eq.~(\ref{Equation}) reduces to a purely
Markovian master equation, ${\partial \rho }/{\partial
  t}=\mathcal{L}\rho(t)$, when $k(t')=\delta(t')$, as expected for a
memoriless channel.

\textit{Dynamical map}.--- We now analytically derive the dynamical map $%
\Phi (t):\rho (0)\mapsto \rho (t)$ governing our master equation. We solve
the integro-differential equation (\ref{Equation}) by taking the Laplace
transform: 
\begin{equation}
s\widetilde{\rho }(s)-\rho (0)=[\widetilde{k}(s)\ast \frac{\mathcal{L}}{ s-%
\mathcal{L}}]\widetilde{\rho }(s),  \label{Lap}
\end{equation}
where $\widetilde{X}(s):=\mathsf{Lap}[X(t)]$ is the Laplace transform of the
function $X(t)$. Now consider the solution of the eigenvalue equation $%
\mathcal{L }\rho =\lambda \rho $. It results in a set of (complex)
eigenvalues $\{\lambda _{i}\}$ and corresponding right and left eigenvectors 
$\{R_{i}\},\{L_{i}\}$ that fulfill the orthonormality condition $\mathrm{Tr}
[L_{i}R_{j}]=\delta _{ij}$. These eigenvectors are known as the damping
basis \cite{Briegel:93} of the superoperator $\mathcal{L}$. Expressing the
density matrix in this basis as $\rho (t)=\sum_{i}\mathrm{Tr}[L_{i}\rho
(t)]R_{i}=\sum_{i}\mu _{i}(t)R_{i}$ and taking the Laplace transform, allows
us to use Eq.~(\ref{Lap}) to solve for the expansion functions $\mu _{i}(t)$%
: $s\widetilde{\mu }_{i}(s)-\mu _{i}(0) =\lambda _{i}\widetilde{k}(s-\lambda
_{i})\widetilde{\mu }_{i}(s)\implies$ 
\begin{eqnarray}
\mu _{i}(t) = \mathsf{Lap}^{-1}[\frac{1}{s-\lambda _{i}\widetilde{k}%
(s-\lambda _{i})} ]\mu _{i}(0)=:\xi _{i}(t)\mu _{i}(0).  \label{eq:Xi}
\end{eqnarray}
The functions $\xi _{i}(t)$ can now be computed using the residue theorem
formula applied to the Bromwich integral formula for the inverse Laplace
transform: if $f(s)=\mathsf{Lap}[F(t)]$ then $%
F(t)=\sum_{p_{k}}\mathrm{Res}[e^{st}f(s),p_{k}]$, where $p_{k}$ are the
poles of $e^{st}f(s)$ and $\mathrm{Res}[g,p]:=\frac{1}{(n-1)!}\left( \frac{
d^{n-1}}{ds^{n-1}}[(s-p)^{n}g(s)]\right) _{s=p}$ is the residue of $g$, with 
$n$ the order of the pole $p$. In our case $f(s)=[s-\lambda _{i}\widetilde{k}
(s-\lambda _{i})]^{-1}$ and so the poles $p_{k}$ are determined by the
solutions of the equation $s=\lambda _{i}\widetilde{k}(s-\lambda _{i})$ for $%
s$. This equation can be solved once the Lindblad generator $\mathcal{L}$
(yielding the $\lambda _{i}$) and the memory kernel $k(t)$ are specified.
Then $\xi _{i}(t)=\sum_{p_{k}^{(i)}}\mathrm{Res}[e^{st}f(s),p_{k}^{(i)}]$.
Summarizing, the dynamical map corresponding to Eq.~(\ref{Equation}) is 
\begin{equation}
\Phi (t):X \mapsto \sum_{i}\xi _{i}(t)\mathrm{Tr}[L_{i}X]R_{i}.  \label{map}
\end{equation}
Using the orthonormality of the damping basis it follows that $\Phi
(t)^{-1}:Y\mapsto \sum_{i}\xi _{i}(t)^{-1}\mathrm{Tr}[L_{i}Y]R_{i}$. Thus $%
\Phi $ is invertible with the exception of the points where $\xi _{i}(t)=0$.
For contractive (e.g., Markovian) maps this will happen at $t=\infty $,
though in general additional points cannot be excluded.

\textit{Condition for complete positivity of} $\Phi$.---
Let us recall
Choi's theorem \cite{Choi:75}: \textquotedblleft Let $\Psi : GL(n;\mathbb{C}%
) \mapsto GL(m;\mathbb{C})$ be a linear map. Then $\Psi $ is CP iff$\ $the
matrix whose elements are $\{\Psi \lbrack E_{ij}]\}_{1\leq i,j\leq n}$ is
positive, where $E_{ij}$ is a matrix with $(E_{ij})_{i,j}=1$ and all other
elements zero.\textquotedblright\
Using Choi's theorem \cite{Choi:75} the criterion for
complete positivity of our map is equivalent to positivity of the matrix $P$
whose $(i,j)$th element is $\Phi \lbrack |i\rangle \langle j|]$. Namely, $%
P\geq 0\Leftrightarrow \{\sum_{k}\xi _{k}(t)\mathrm{Tr}[L_{k}|i\rangle
\langle j|]R_{k}\}_{1\leq i,j\leq n}=\{\sum_{k}\xi _{k}(t)\langle
j|L_{k}|i\rangle R_{k}\}_{1\leq i,j\leq n}\geq 0$, which, in turn, is
equivalent to: 
\begin{equation}
\sum_{k}\xi _{k}(t)L_{k}^{T}\otimes R_{k}\geq 0.  \label{eq:CPtest}
\end{equation}
The inequality~(\ref{eq:CPtest}) is a necessary and sufficient condition for
our map to be CP. Because the functions $\xi _{k}(t)$ are given in terms of
the memory kernel $k(t)$ through Eq.~(\ref{eq:Xi}), this inequality results
in a condition on $k(t)$, which can be checked in order to verify that a
given such kernel results in a CP map. Further note that Eq.~(\ref{Equation}%
) preserves the trace of $\rho (t)$ [i.e., $d\mathrm{\ Tr}\rho (t)/dt=0$],
as is evident from $\mathrm{Tr}\mathcal{L}=0$ and a Taylor expansion of $%
\exp (\mathcal{L}t)$.

\textit{Kraus representation of} $\Phi $.--- Since the matrix $P$ is
positive it can be expressed as $P=\sum_{k}|a_{k}\rangle \langle a_{k}|$
where the $|a_{k}\rangle $'s are the eigenvectors of $P$. One can divide the
vector $|a_{k}\rangle $ into $n$ segments of length $n$, where $n= \dim[%
\mathcal{H}_{S}]$, and define a matrix $M_{k}$ with the $i$th column being
the $i$th segment of $|a_{k}\rangle $, so that the $i$th segment is $%
M_{k}|i\rangle $. Then the dynamical map is reconstructed as $\mathcal{E}%
(\rho)=\sum_{\alpha }M_{\alpha }\rho M_{\alpha }^{\dag }$, which is the
desired Kraus representation.

\textit{Connection to other master equations}.--- We first note that our
master equation (\ref{Equation}) is an instance of the exact
Nakajima-Zwanzig (NZ) equation $\overset{.}{\rho}=\int_{0}^{t}dt^{ \prime
}O(t,t^{\prime })\rho(t^{\prime })$ \cite{Nakajima:58Zwanzig:60a}, where the
NZ kernel $O(t,t^{\prime })$ is, in our case, of the special time
translationally-invariant form $O(t^{\prime }-t)$. Secondly, in the
particular case that $||\mathcal{L}||\ll 1/t $ Eq.~(\ref{Equation}) reduces
to 
\begin{equation}
\frac{\partial \rho}{\partial t}=\mathcal{L}\int_{0}^{t}dt^{\prime
}k(t^{\prime })\rho(t-t^{\prime }).  \label{eq:intui}
\end{equation}
This master equation was proposed intuitively in Ref.~\cite{barnett:033808},
where it was studied in the case of a damped harmonic oscillator and it was
shown to lead, under certain assumptions, to unphysical behavior. This issue
was clarified in the recent work \cite{Daffer:03}, where it was shown that a
single qubit subject to telegraph noise can be described by Eq.~(\ref%
{eq:intui}), and where conditions for complete positivity of (\ref{eq:intui}%
) were established; our inequality (\ref{eq:CPtest}) includes this as a
special case. Thirdly, we can rewrite Eq.~(\ref{Equation}) in
time-convolutionless form using the backward propagator method \cite%
{Shibata:77Chaturvedi:79}: Using Eq.~(\ref{map}) we can express the formal
solution of Eq.~(\ref{Equation}) as $\rho(t)=\Phi (t)\rho(0)$. We have
already discussed above the invertibility of $\Phi (t)$; assuming $\Phi
^{-1} $ exists Eq.~(\ref{Equation}) can then be rewritten in
time-convolutionless form as 
\begin{equation}
\frac{\partial \rho}{\partial t}=\left[ \mathcal{L}\int_{0}^{t}k(t^{\prime
})\exp (\mathcal{L}t^{\prime })\Phi (t-t^{\prime }) dt^{ \prime } \Phi
^{-1}(t) \right] \rho(t),  \label{eq:TCL}
\end{equation}
with the operator in square brackets serving as the generator of the
evolution.
%%Finally, we wish to address the problem of unravelling the
%%dynamics described by Eq.~(\ref{Equation}) into quantum trajectories \cite%
%%{Imamoglu:94,Breuer:04,Dalibard:92Gisin:92Plenio:98,gambetta:012108breuer:022115}%
%%. Since Eq.~(\ref{eq:TCL}) is convolutionless and CP-preserving, it can be
%%transformed into a Lindblad equation [Eq.~(\ref{eq:Lind})] with \emph{%
%%time-dependent} coefficients $a_{\alpha }(t)$ (see, e.g., \cite{Lidar:CP01}
%%for a method for doing this starting from the Kraus operator sum
%%representation). There then exists a continuous bath-measurement
%%interpretation, and consequently, an unravelling of the dynamics into
%%quantum trajectories, as shown explicitly, e.g., in Ref.~\cite{Breuer:04}.

\textit{Experimental determination of the kernel function}.--- Suppose one
measures $\rho (t)$ via quantum state tomography (QST) \cite{Nielsen:book}.
It then follows from Eq.~(\ref{map}) applied to $\rho (t)$ that $\xi _{i}(t)=%
\mathrm{Tr}[L_{i}\rho (t)]/\mathrm{Tr} [L_{i}\rho (0)]$. The coefficients $%
\xi _{i}(t)$ are thus directly experimentally accessible, \emph{provided one
first specifies a Markovian model} from which the left eigenvectors $L_{i}$
and eigenvalues $\lambda _{i} $ can be computed. Inverting Eq.~(\ref{eq:Xi})
then yields the kernel as $k(t)=\mathsf{Lap}^{-1}[(s-1/\mathsf{Lap}[\xi
_{i}(t)])]e^{-\lambda _{i}t}/\lambda _{i}$. This inversion process for $k(t)$
is not unique in the sense that it will depend on the choice of Markovian
model. It can be optimized via well-established maximum likelihood methods,
e.g., \cite{Banaszek:99}, thus yielding the \emph{optimal} Markovian model.

\textit{Example}.--- As a concrete example meant to illustrate the
predictions of our master equation we consider the problem of a single qubit
dephasing. The Lindblad superoperator is $\mathcal{L}\rho =-(a/2)[\sigma
_{z},[\sigma _{z},\rho ]]$, $a>0$. Using the parametrization $\rho (t)=(I+%
\vec{\alpha}(t)\cdot \vec{\sigma})/2$ [with $\vec{\alpha }\in \mathbb{R}^{3}$
and $\vec{\sigma}=(\sigma _{x},\sigma _{y},\sigma _{z})$], the damping basis
is found to consist of the following eigenvalues and eigenoperators: $%
\{\lambda _{i}\}_{i=0}^{3}=\{0,-a,-a,0\}$, and $\{R_{i}\}_{i=0}^{3}=\{L_{i}%
\}_{i=0}^{3}=\{I,\sigma _{x},\sigma _{y},\sigma _{z}\}/\sqrt{2}$. The
Markovian solution is simple exponential coherence decay:\ $\alpha _{z}(t)=1$
and $\alpha _{j}(t)=\alpha _{j}(0)\exp (-at)$, $j=x,y$. It follows
immediately from Eq.~(\ref{eq:Xi}) that $\xi _{0}(t)=\xi _{z}(t)=\mathsf{Lap}%
^{-1}[1/s]=1$ and that $\xi _{x}(t)=\xi _{y}(t)=:f(t)$. We further find $%
\{\alpha _{j}(t)=\xi _{j}(t)\}_{j=x,y,z}$. Applying the criterion (\ref%
{eq:CPtest}) readily yields the CP condition as $|f(t)|\leq 1$. Let us
consider two kernel functions: $k_{1}(t)=A\exp (-\gamma t)\Rightarrow 
\widetilde{k}_{1}(s)=\frac{ A}{s+g}$ and $k_{2}(t)=Ae^{-(\gamma -a)t}[\cos
(\mu t)-\frac{\gamma }{\mu } \sin (\mu t)]\Rightarrow \widetilde{k}_{2}(s)=%
\frac{A(s-a)}{(s-a+\gamma )^{2}+\mu ^{2}}$. Then, following the prescription
of Eq.~(\ref{eq:Xi}) yields $f_{1}(t)=\exp [-t(a+\gamma )/2][\cos (\omega
t)+\sin (\omega t)(a+\gamma )/2\omega ]$ where $\omega =\sqrt{4Aa-(\gamma
+a)^{2}}/2$, and $f_{2}(t)=1-\frac{Aa}{\gamma ^{2}+\Omega ^{2}}[1-e^{-\gamma
t}(\cos \Omega t+ \frac{\gamma }{\Omega }\sin \Omega t)]$ where $\Omega =%
\sqrt{\mu ^{2}+Aa}$ (note that the CP condition $|f_{1,2}(t)|\leq 1$ imposes
restrictions on the allowed values of the various parameters appearing
here). In both cases we thus find damped oscillations. The difference is
that in the case of $k_{1}$ we have $f_{1}(\infty )=0$, as in the Markovian
case, while in the case of $k_{2}$ we have $f_{2}(\infty )=$ $1-\frac{Aa}{%
\gamma ^{2}+\Omega ^{2}}$, which cannot be mimicked by the Markovian
solution. Damped oscillations with a non-zero asymptotic coherence, as in
the case of $k_{2}$, are a feature of the exact solution of a single qubit
dephasing in the presence of a boson bath, e.g., when a peaked spectral
density $g(\omega )\propto \exp [-c(\omega -\omega _{0})^{2}]$ is chosen 
\cite{Lidar:CP01}. We thus see explicitly through the example considered
here, how our new master equation (\ref{Equation}) is capable of
interpolating between exact and Markovian open system dynamics.

Financial support from the Sloan Foundation and the DARPA-QuIST program
(managed by AFOSR under agreement No. F49620-01-1-0468) is gratefully
acknowledged (to D.A.L).

\end{document}